\documentclass[referee]{raa}            

\usepackage{graphicx,times}             
\usepackage{natbib}
\usepackage{amssymb,amsmath}
\usepackage{longtable}
\usepackage{threeparttable}
\usepackage{multirow}
\bibpunct{(}{)}{;}{a}{}{,}

\usepackage[pagebackref=true]{hyperref}

\begin{document}

  \title{Preliminary Exploration of Areal Density of Angular Momentum for Spiral Galaxies}

   \volnopage{Vol.0 (20xx) No.0, 000--000}      
   \setcounter{page}{1}          

   \author{Lan Zhang 
      \inst{1}
   \and Feilu Wang
      \inst{1, 2}
   \and Xiangxiang Xue
      \inst{1, 3}
    \and David Salzmann
      \inst{1}
    \and Baifei Shen
      \inst{4}
    \and Zehao Zhong
      \inst{1}
    \and Gang Zhao
      \inst{1, 2}               
   }

   \institute{CAS Key Laboratory of Optical Astronomy, National Astronomical Observatories, Chinese Academy of Sciences, Beijing, 100101, People’s Republic of China; {\it wfl@bao.ac.cn} \\
        \and
             School of Astronomy and Space Science, University of Chinese Academy of Sciences, Beijing, 101408, People’s Republic of China\\
        \and
             Institute for Frontiers in Astronomy and Astrophysics, Beijing Normal University, Beijing, 102206, People’s Republic of China\\
         \and
             Department of Physics, Shanghai Normal University, Shanghai, 200234, China\\
\vs\no
   {\small Received 20xx month day; accepted 20xx month day}}

\abstract{The specific angular momenta ($j_t$) of stars, baryons as a whole and dark matter haloes contain clues of vital importance about how galaxies form and evolve. Using a sample of 70 spiral galaxies, we perform a preliminary analysis of $j_t$, and introduce a new quantity, e.g., areal density of angular momentum (ADAM) ($j_t~M_\star/4R_d^2$) as an indication for the existence of jet in spiral galaxies. The percentage of spiral galaxies having jet(s) shows strong correlation with the ADAM, although the present sample is incomplete.
\keywords{galaxies: spiral --- galaxies: jets}
}
   \authorrunning{L. Zhang, F.L. Wang et al.}            
   \titlerunning{Jet - Angular Momentum Coupling in the Spiral Galaxies}  

   \maketitle

%
%
\section{Introduction}
\label{sec:intro}

The first evidence of jet-like features emanating from the nuclei of galaxies goes back to the discovery by \cite{1918PLicO..13....9C}  of the optical jet from the elliptical galaxy NGC 4486 (M87) in the Virgo cluster. According to the definition of \cite{1984ARAA..22..319B}  for extragalactic jets, the term ``jets" is used to designate collimated ejecta that have opening angles $\leqslant 15^{\circ}$.

Jets of matter occur in many astrophysical situations, but can broadly be classified into two types: stellar jets and galactic jets. Stellar jets are generated by a number of different sources, such as T Tauri stars, planetary nebula, neutron stars, and stellar black holes. Galactic jets, however, are believed to have only a single source, namely, a supermassive black hole (SMBH) at the centre of the galaxy. 

Many attempts have been made to understand the formation of galactic jets by testing various relations among the physical and observational properties of the host galaxy and its central SMBH. Several options for such relations have been suggested in the literature,  \\
 (1) correlation between jet properties and central black hole mass/spin/accretion rate, e.g., \cite{2000apjLaor, Dotti2013ApJ, Narayan1995ApJ};\\
 (2) correlation between jet properties and accretion disk spin, e.g., \cite{Natarajan1998,2001ApJ...555..650H,2013ApJ...762..103K}; \\
 (3) correlation between jet formation and magnetic fields, e.g., \cite{Asada2002pasj,2008PPCF...50l4057T} . 
 
Fully evolved relativistic jets have traditionally been associated with high-mass elliptical galaxies hosting the most massive black holes, but \cite{2022arXiv220400020} confirmed that also less massive black holes in spiral galaxies could launch and sustain powerful jets, implying that the launching of the jets is governed by factors other than those mentioned previously.

The role of the angular momentum of galaxies plays in the jet properties is little understood, although it is believed to control the kinematics of their stars, which on the other hand drives observable quantities such as the apparent radius, the bulge fraction, and the alignment with other nearby structures \citep{cima20}. The galactic angular momenta originate from the initial spin, and lose during the mergers. It plays a major role in galaxy formation and evolution and is closely related to the coupling between dark and visible matter \citep{li2022}. The primary goal of this paper is to study the correlation between the angular momenta and the presence of jet in the spiral galaxies embedding an SMBH. 

\section{$\lowercase{j_t}-M_\star$ diagram for spiral galaxies}
\label{sec:jt_m}

Our sample consists of 67 spiral galaxies in Table 4 of \cite{Roman12}, whose bulge-to-total mass ratios $B/T$ range from $0.0$ to $0.6$, that correspond to Sc (pure disks) to S0, and three spiral galaxies, in order to extend the stellar mass range of samples in this study. These are the MilkyWay and 2MASX
J23453268-0449256 (hereafter J2345-0449) with large stellar mass, and NGC 4395 with small stellar mass were added to construct our sample.

Following \cite{Roman12}, the total specific stellar angular momenta ($j_t$, in the unit of ${\rm km~s^{-1}~kpc}$) of all the samples are calculated by
\begin{equation}
\label{eq:jt}
    j_t = f_b j_b + (1 - f_b) j_d,
\end{equation}
where $f_b$ is the bulge stellar mass fraction, $j_b$ and $j_d$ are intrinsic values of specific stellar angular momentum of bulge and disk components, respectively. Both $j_b$ and $j_d$ are calculated from the values along the projected semimajor axis of galaxies ($j_{p, b}$ and $j_{p, d}$).
For the bulge part, 
\begin{equation}
\label{eq:jp}
    j_{p, b} = k_n v_{s, b} a_{e, b},
\end{equation}
where $v_{s, b}$ is the observed rotation velocity of bulge, $k_n \sim 1-5$ is a numerical coefficient that depends on the Sersic index $n$ of the galaxy (Eq. A31 in \cite{Roman12}) and $a_{e, b}$ is the effective radius along the semimajor axis. 
While for the disk component,
\begin{equation}
\label{eq:jt_ap}
    j_{p, d} = 2 v_{c} \sin i R_d,
\end{equation}
where $R_d$ is the intrinsic exponential-disk scale length, $v_{c}$ is the intrinsic circular rotation velocity of disk, based on the rotation curves over the range $(2 - 3) R_d$, and $i$ is the inclination. Then the intrinsic values of $j$ is converted by the deprojection factor $C_i$ which correlated with $i$, 
\begin{equation}
\label{eq:jp_t}
    j = j_p C_i
\end{equation}
For the bulge component.
\begin{equation}
\label{eq:ci_b}
    C_i \simeq \frac{0.99 + 0.14 i}{\sin i}
\end{equation}
and for disk component, 
\begin{equation}
\label{eq:ci_d}
    C_i = \frac{1}{\sin i}
\end{equation}

For the Milky Way, NGC 4395, and the giant radio source J2345-0449, their $j_t$ are calculated by using Eq.~\ref{eq:jt} and observable data from literature studies, which listed in Tab.~\ref{tab:data}.

The stellar specific angular momentum-mass relation was first studied by \cite{1983IAUS..100..391F}. Follow-up studies confirmed the relation with more and better data (e.g. \cite{Roman12, 2018ApJ...868..133F, 2018A&A...612L...6P}). However, no previous literature studies explore that whether the jets may effect the relation or nor. For better understanding the role of jets in the stellar specific angular momentum-mass relation, especially for spiral galaxies with jets, we modeled $j_t$ ($j_{t, {\rm mod}}$) as a function of $M_{\star}$ in $\log-\log$ space for samples with/without jets, respectively:
\begin{equation}
\label{eq:jt_mod}
\log j_{t, {\rm mod}} = \beta + \alpha \times \log\left(\frac{M_{\star}}{M_\odot}\right),
\end{equation}
We carried out the Markov-Chain Monte-Carlo (MCMC) to explore the relation and the uncertainties of the fitted parameters $\alpha$ and $\beta$, for three groups of our sample, that is, a) all the sample galaxies; b) all possible jetted galaxies; c) all possible jetted galaxies excluding NGC 3898, NGC 4258, NGC 4736, and NGC 5033, because the jets of these four galaxies are still in doubt \citep{Baldi_2021}. 

\begin{figure}[htp]
  \centering
  \includegraphics[width=0.75\linewidth]{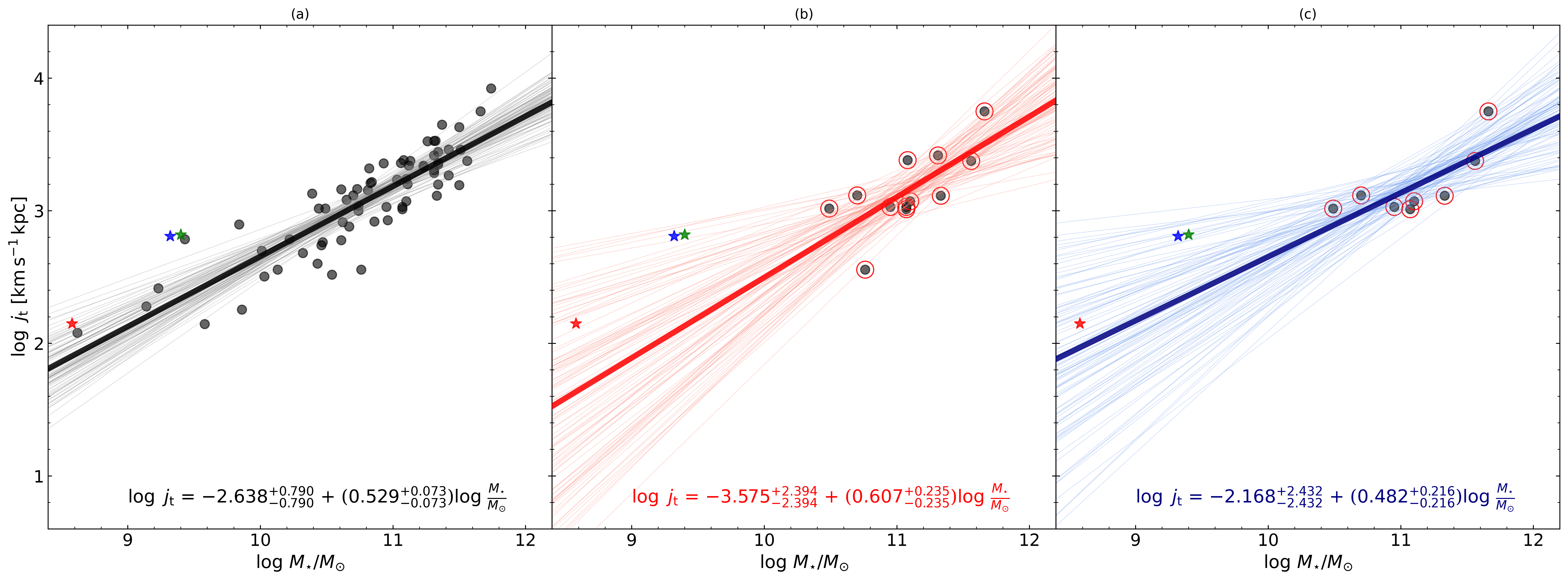}
  \caption{The total angular momenta of our sample versus their stellar mass. For all plots, black dots are observation data, while galaxies with jets are marked with red circles. Thick lines are best fits of $\log j_{\rm t} - \log \frac{M_{\star}}{M_{\odot}}$ relations. Star symbols indicate NGC 4395, which was excluded during the fittings. Red, blue, and green represent that its $j_t$ is calculated with stellar mass and velocity, gas mass and velocity, and total baryonic mass and velocity, respectively.The shade of each line shows $1\sigma$ fitting error range. a) shows the fitting results of all spiral galaxies in the present study; b) shows the fittings for all possible spiral galaxies with jets; c) is similar as b), but with the assumption that NGC 3898, NGC 4358, NGC 4736, and NGC 5033 are spiral galaxies without jets.}
  \label{fig:1}
\end{figure}

The fitting also does not include NGC 4395 since its mass is much lower than the others and is an irregular galaxy whose gas component seem to dominate \citep{Repetto2017}. Our fitting results for three groups are shown in Fig.~\ref{fig:1} and Tab.~\ref{tab:fit}.

Here the fitting parameters for all sample are $\alpha=0.529^{+0.073}_{-0.073}$ and $\beta=-2.638^{+0.790}_{-0.790}$, which show good agreement with the study of \cite{Roman12}, where $\alpha=0.52\pm0.04$ and $\beta=-2.54\pm0.05$ for all spirals in their work.
While for the group with jets, $\alpha=0.607^{+0.235}_{-0.235}$ and $\beta = -3.575^{+2.394}_{-2.394}$, which show large fitting uncertainties. The reason is that this group is concentrated in the large stellar mass range. For NGC 4395, if its $j_t$ is calculated with the stellar mass and stellar velocity, it shows good agreement with the fitted relation. If all baryonic matters are included, the $j_t$ deviates from the best fitting, although it is still covered by $1\sigma$ fitting uncertainty. Unlike stellar composition, the gas component is not in an equilibrium 
state, its high velocity may be speeded up by the jet, which results in a large $j_t$ value.

\section{Percentage of galaxies with observed jet(s) }
After cross-matching the samples with recent radio- and x-ray observation studies \citep[e.g., ][]{2013ApJ...774L..25K, Baldi_2021}, 13 galaxies are labeled as ones with jets. Among them, there are nine galaxies with the presence of a jet or jets in the observations, i.e., NGC 2639 \citep{2019MNRAS.490L..26S}, NGC 3031 (M81)\citep{2019MNRAS.488.4317B}, NGC 4395\citep{2013ApJ...774L..25K},  NGC 4594 (M104)\citep{2013ApJ...779....6H}, NGC 2841\citep{Baldi_2018}, NGC 3198\citep{Baldi_2018},  NGC 7217\citep{Baldi_2018}, J2345-0449 \citep{Bagchi_2014} and the Milky Way\citep{2021ApJ...922..254C}. For the rest four galaxies, NGC 4258 (M106) was identified as a galaxy with jet in \cite{2000ApJ...536..675C}, however, in the recent study of \cite{Baldi_2021}, the jet morphs of NGC 4258 (M106), as well as NGC 3898, NGC 4736, and NGC 5033, were uncertain. Therefore, NGC3898, NGC 4258, NGC 4736, and NGC 5033 may not be the galaxies with jets \citep{Baldi_2021}.

 
Fig. 2 gives the percentage of galaxies with jet(s) in our sample to find a better indication for the existence of jet in spiral galaxies. Here we introduce a new quantity, e.g., the areal density of angular momentum (ADAM), which is defined by $j_t~M_\star/(R_{\rm eff})^2$, where $R_{\rm eff}$ is the effective radius of the stellar mass. For the present samples, we take $2R_d$ as the effective radius. Therefore, the ADAM adopted in this work is $j_t~M_\star/(2R_d)^2$. In order to ensure the percentage is statistically significant, the number of samples in each $j_t$ or ADAM bin is more than or equal to 3.
Compared $j_t$ with ADAM,
the latter seems to be a better indicator of the presence of jet, since the percentage is lower than $20\%$ when ADAM is less than $~10^{12}~{\rm km/s}\cdot M_\odot$ kpc.
What makes $j_t$ a quantity of great astrophysical importance is not only its relation to the baryonic matter content of galaxies, but also its connection with galactic jet(s).

The present results are preliminary. First, our samples are limited to spiral galaxies. As far as we know, more jets launching in elliptical galaxies. However, spiral galaxies have, on the average, higher angular momentum than elliptical ones for a given stellar mass \citep{2022arXiv220611913C,1983IAUS..100..391F}. Secondly, our galaxy sample is neither statistically complete nor the average distribution of the parameters, e.g., stellar mass, BH mass and $j_t$. In addition, 
the scale of jets is not taken into account, which is from several pc to $10^2 - 10^3$~kpc. 
It may matter, since the relation of accretion mass with the jet luminosity seems to be valid for kpc-scale jets, but not for pc-scale \citep{2013MNRAS.432.1138P}.

As to the possible relation between the ADAM and the jet, it is well known that tornado can suck matters on earth and solar storm is related to magnetic helicity. We carried a numerical simulation by putting plasma in an axial magnetic field and a radial electrical field which could be formed by quick shrink of the galaxies. In real case the electrical field could be replaced with gravitational force. In simulation, we find the particles rotate around the axis and jets emitted along the axis \citep{2018LPB....36..384Q}. The rotation of the particles can also be driven by the twist laser pulse. In that case, jets were also observed in simulation \citep{2019PhRvL.122b4801W}.

\begin{figure}[htp]
  \centering
  \includegraphics[width=0.55\linewidth]{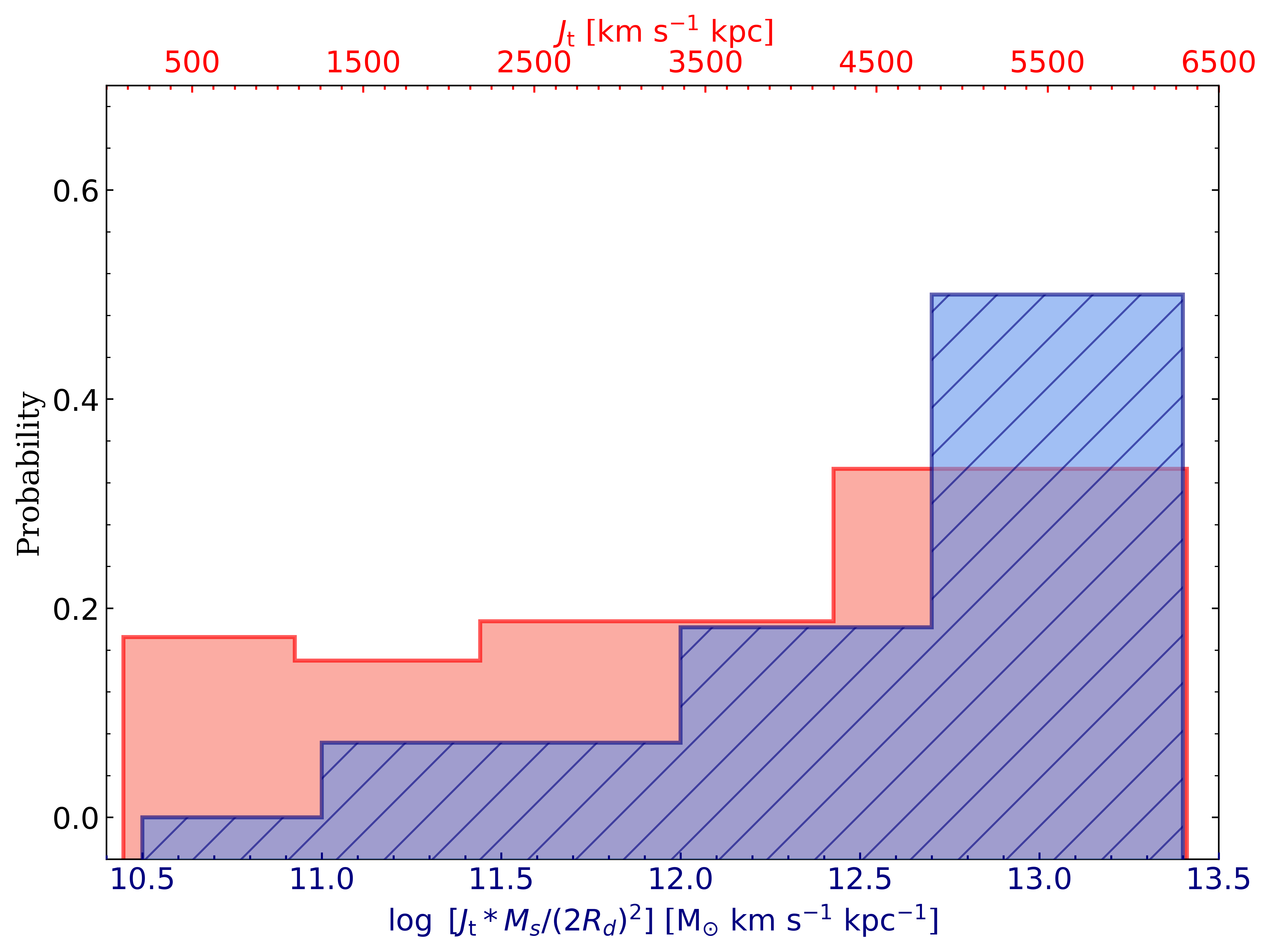}
  \caption{The percentage of galaxies having jet(s) in the sample as function of the specific angular momenta $j_t$ (in red, upper scale) and those having areal density of angular momentum (ADAM) $j_t~M_\star/(2R_d)^2$ (in blue, lower scale).}
\end{figure}

\begin{table}[htp]
\begin{threeparttable}
\caption{Observations used in $j_t$ calculation}
\centering
\begin{tabular}{lccccccclc}
\hline
Name & $R_d$  & $f_b$ & $a_{e, b}$ & $i$    & $v_{s, b}$         & $v_{c, d}$       & $j_t$                    &  $\log (M_\star/M_\odot)$  &  $\log (M_{BH}/M_\odot)$  \\
         &	(kpc)  &         & (kpc)         & (deg) &  (km~s$^{-1}$)  & (km~s$^{-1}$) & (km~s$^{-1}$~kpc) &                                  	  &                                         \\
\hline
Milky Way     & 3.00$^a$ & 0.00 & 0.00 & ...     & ... & 218$^a$ & 1308 &  10.70 & 6.63$^b$ \\
\hline
\multirow{3}{*}{NGC 4395} & \multirow{3}{*}{3.93$^c$} & \multirow{3}{*}{0.00} & \multirow{3}{*}{0.00} & \multirow{3}{*}{46.9$^c$} & \multirow{3}{*}{...} & 18$^{c, d}$& 141$^{d}$ & 8.58$^{c, d}$& \multirow{3}{*}{5.64$^g$} \\
\multirow{3}{*}{} & \multirow{3}{*}{} & \multirow{3}{*}{} & \multirow{3}{*}{} & \multirow{3}{*}{} & \multirow{3}{*}{} & 83$^{c, e}$ & 644$^e$ & 9.32$^{c, e}$ & \multirow{3}{*}{} \\
\multirow{3}{*}{} & \multirow{3}{*}{} & \multirow{3}{*}{} & \multirow{3}{*}{} & \multirow{3}{*}{} & \multirow{3}{*}{} & 85$^{c, f}$  & 660$^f$ & 9.40$^{c, f}$ & \multirow{3}{*}{} \\
\hline
J2345-0449 & 7.51$^h$  & 0.15$^h$ & 1.18$^h$  & 59.7$^h$  & 266$^{h, i}$ & 371$^h$ & 5625 & 11.66$^h$ & $\ge$8.30$^j$\\
\hline
\end{tabular}
\begin{tablenotes}
\tiny
\item Notes. References are $a-$\cite{Bovy2012}; $b-$\cite{Marasco2021}; $c-$\cite{Repetto2017}; $d-$ gas only; $e-$ star only; $f-$ total baryonic matters;
$g-$\cite{2018ApJ...869..113D};$h-$ \cite{Bagchi_2014}; $i-$ calculated by Eq.~8 of \citet{Bagchi_2014}; $j-$\cite{2021MNRAS.500.2503M};
\end{tablenotes}
\label{tab:data}
\end{threeparttable}
\end{table}

\begin{table}[htp]
\caption{Fits to stellar mass and specific angular momenta}
\centering
\begin{tabular}{lcc}
\hline
Sample & $\alpha$ & $\beta$ \\ 
\hline
All spiral galaxies & $0.529^{+0.073}_{-0.073}$ & $-2.638^{+0.790}_{-0.790}$ \\
spiral galaxies with jets & $0.607^{+0.235}_{-0.235}$ & $-3.575^{+2.394}_{-2.394}$ \\
spiral galaxies without doubt jets & $0.482^{+0.216}_{-0.216}$ & $-2.168^{+2.432}_{-2.432}$ \\
\hline
\end{tabular}
\label{tab:fit}
\end{table}

\normalem
\begin{acknowledgements}
We thank Dr. Lin Zhu, \& Dr. Dawei Xu for useful suggestions and discussions, and the anonymous referee for helpful comments.
This work is supported by National Natural Science Foundation of China (NSFC) under grants No. 11988101, 12073043, and National
Key Research and Development Program of China No. 2019YFA0405500.
L.Z. and X-X.X. acknowledge the support from CAS Project for Young Scientists in Basic Research Grant No. YSBR-062.

\end{acknowledgements}

\bibliography{mybib}
\bibliographystyle{raa}

\label{lastpage}

\end{document}